\PassOptionsToPackage{dvipsnames,svgnames}{xcolor}
\documentclass[10pt,a4paper,onecolumn]{article}

\usepackage{marginnote}    
\usepackage{graphicx}      
\usepackage{xcolor}        
\usepackage{etoolbox}      
\usepackage[affil-it]{authblk} 
\usepackage{orcidlink}     
\usepackage{titlesec}      
\usepackage{calc}          
\usepackage{tikz}          
\usepackage{hyperref}      
\usepackage{caption}       
\usepackage{amssymb,amsmath} 
\usepackage{ifxetex,ifluatex} 
\usepackage{seqsplit}      
\usepackage{xstring}       
\usepackage{float}         

\let\origfigure\figure
\let\endorigfigure\endfigure
\renewenvironment{figure}[1][2] {
  \expandafter\origfigure\expandafter[H]
} {
  \endorigfigure
}

\NewDocumentCommand\citeproctext{}{}
\NewDocumentCommand\citeproc{mm}{%
  \begingroup\def\citeproctext{#2}\cite{#1}\endgroup}
\makeatletter
 \let\@cite@ofmt\@firstofone
 \def\@biblabel#1{}
 \def\@cite#1#2{{#1\if@tempswa , #2\fi}}
\makeatother

\newlength{\cslhangindent}
\newlength{\csllabelwidth}
\newenvironment{CSLReferences}[2] 
 {\begin{list}{}{%
  \setlength{\itemindent}{0pt}
  \setlength{\leftmargin}{0pt}
  \setlength{\parsep}{0pt}
  \ifodd #1
   \setlength{\leftmargin}{\cslhangindent}
   \setlength{\itemindent}{-1\cslhangindent}
  \fi
  \setlength{\itemsep}{#2\baselineskip}}}
 {\end{list}}

\ifLuaTeX
  \usepackage[bidi=basic]{babel}
\else
  \usepackage[bidi=default]{babel}
\fi
\babelprovide[main,import]{american}

\DeclareUnicodeCharacter{2265}{\ensuremath{\geq}}   
\DeclareUnicodeCharacter{0308}{\"{}}

\let\textttOrig=\texttt
\def\texttt#1{\expandafter\textttOrig{\seqsplit{#1}}}
\renewcommand{\seqinsert}{\ifmmode
  \allowbreak
  \else\penalty6000\hspace{0pt plus 0.02em}\fi}

\makeatletter
\let\href@Orig=\href
\def\href@Urllike#1#2{\href@Orig{#1}{\begingroup
    \def\Url@String{#2}\Url@FormatString
    \endgroup}}
\def\href@Notdoi#1#2{\def\tempa{#1}\def\tempb{#2}%
  \ifx\tempa\tempb\relax\href@Urllike{#1}{#2}\else
  \href@Orig{#1}{#2}\fi}
\def\href#1#2{%
  \IfBeginWith{#1}{https://doi.org}%
  {\href@Urllike{#1}{#2}}{\href@Notdoi{#1}{#2}}}
\makeatother

\usepackage[top=3.5cm, bottom=3cm, right=1.5cm, left=1.0cm,
            headheight=2.2cm, reversemp, includemp, marginparwidth=4.5cm]{geometry}

\titleformat{\section}
  {\normalfont\sffamily\Large\bfseries}
  {}{0pt}{}
\titleformat{\subsection}
  {\normalfont\sffamily\large\bfseries}
  {}{0pt}{}
\titleformat{\subsubsection}
  {\normalfont\sffamily\bfseries}
  {}{0pt}{}
\titleformat*{\paragraph}
  {\sffamily\normalsize}

\usepackage{fancyhdr}
\makeatletter
\let\ps@plain\ps@fancy
\fancyheadoffset[L]{4.5cm}
\fancyfootoffset[L]{4.5cm}
\definecolor{linky}{rgb}{0.0, 0.5, 1.0}

\newcommand{\ExternalLink}{%
   \tikz[x=1.2ex, y=1.2ex, baseline=-0.05ex]{%
     \begin{scope}[x=1ex, y=1ex]
       \clip (-0.1,-0.1)
         --++ (-0, 1.2)
         --++ (0.6, 0)
         --++ (0, -0.6)
         --++ (0.6, 0)
         --++ (0, -1);
       \path[draw,
         line width = 0.5,
         rounded corners=0.5]
         (0,0) rectangle (1,1);
     \end{scope}
     \path[draw, line width = 0.5] (0.5, 0.5)
       -- (1, 1);
     \path[draw, line width = 0.5] (0.6, 1)
       -- (1, 1) -- (1, 0.6);
     }
   }

\patchcmd{\@maketitle}{center}{flushleft}{}{}
\patchcmd{\@maketitle}{center}{flushleft}{}{}
\patchcmd{\@maketitle}{\LARGE}{\LARGE\sffamily}{}{}
\def\maketitle{{%
  
  \AB@maketitle}}
\makeatletter
\renewcommand\AB@affilsepx{ \protect\Affilfont}
\renewcommand\AB@affilnote[1]{{\bfseries #1}\hspace{3pt}}
\renewcommand{\affil}[2][]%
   {\newaffiltrue\let\AB@blk@and\AB@pand
      \if\relax#1\relax\def\AB@note{\AB@thenote}\else\def\AB@note{#1}%
        \setcounter{Maxaffil}{0}\fi
        \begingroup
        \let\href=\href@Orig
        \let\texttt=\textttOrig
        \let\protect\@unexpandable@protect
        \def\thanks{\protect\thanks}\def\footnote{\protect\footnote}%
        \@temptokena=\expandafter{\AB@authors}%
        {\def\\{\protect\\\protect\Affilfont}\xdef\AB@temp{#2}}%
         \xdef\AB@authors{\the\@temptokena\AB@las\AB@au@str
         \protect\\[\affilsep]\protect\Affilfont\AB@temp}%
         \gdef\AB@las{}\gdef\AB@au@str{}%
        {\def\\{, \ignorespaces}\xdef\AB@temp{#2}}%
        \@temptokena=\expandafter{\AB@affillist}%
        \xdef\AB@affillist{\the\@temptokena \AB@affilsep
          \AB@affilnote{\AB@note}\protect\Affilfont\AB@temp}%
      \endgroup
       \let\AB@affilsep\AB@affilsepx
}
\makeatother

\renewcommand\Affilfont{\sffamily\small\mdseries}
\usepackage[T1]{fontenc}
\usepackage[utf8]{inputenc}

\IfFileExists{upquote.sty}{\usepackage{upquote}}{}
\IfFileExists{microtype.sty}{%
\usepackage{microtype}
\UseMicrotypeSet[protrusion]{basicmath} 
}{}

\usepackage{hyperref}
\hypersetup{unicode=true,
            pdfborder={0 0 0},
            breaklinks=true,
            pdflang={en-UK},
            colorlinks=true,
            linkcolor={Maroon},
            filecolor={Maroon},
            citecolor=[rgb]{0.0, 0.0, 1.0},
            urlcolor=linky,
            colorlinks,breaklinks=true,
            urlcolor=linky,
            linkcolor=linky}
\let\addcontentslineOrig=\addcontentsline
\def\addcontentsline#1#2#3{\bgroup
  \let\texttt=\textttOrig\addcontentslineOrig{#1}{#2}{#3}\egroup}
\let\markbothOrig\markboth
\def\markboth#1#2{\bgroup
  \let\texttt=\textttOrig\markbothOrig{#1}{#2}\egroup}
\let\markrightOrig\markright
\def\markright#1{\bgroup
  \let\texttt=\textttOrig\markrightOrig{#1}\egroup}
  
\IfFileExists{parskip.sty}{%
\usepackage{parskip}
}{
\setlength{\parindent}{0pt}
\setlength{\parskip}{6pt plus 2pt minus 1pt}
}
\ifx\paragraph\undefined\else
\let\oldparagraph\paragraph
\renewcommand{\paragraph}[1]{\oldparagraph{#1}\mbox{}}
\fi
\ifx\subparagraph\undefined\else
\let\oldsubparagraph\subparagraph
\renewcommand{\subparagraph}[1]{\oldsubparagraph{#1}\mbox{}}
\fi

\title{aurel: A Python package for automatic relativistic calculations}

\author[1]{Robyn L. Munoz%
    \,\orcidlink{0000-0003-3345-8520}\,%
    }
\author[1]{Christian T. Byrnes%
    \,\orcidlink{0000-0003-2583-6536}\,%
    }
\author[1]{Will J. Roper%
    \,\orcidlink{0000-0002-3257-8806}\,%
    }

\affil[1]{Department of Physics and Astronomy, University of Sussex, Brighton, BN1 9QH, United Kingdom}

\hypersetup{pdftitle={Code paper name},
            pdfauthor={Author1 Name, Author2 Name}}

\begin{document}
\date{}
\maketitle

\marginpar{
  \begin{flushleft}
  \sffamily\small



  {\bfseries Software}
  \begin{itemize}
    \setlength\itemsep{0em}
    \item \href{https://github.com/robynlm/aurel}{\color{linky}{Repository}} \ExternalLink
  \end{itemize}

  \vspace{2mm}
  \par\noindent\hrulefill\par
  \vspace{2mm}


  {\bfseries Published:} unpublished

  \vspace{2mm}
  {\bfseries License}\\
  Authors of papers retain copyright and release the work under a Creative Commons Attribution 4.0 International License (\href{https://creativecommons.org/licenses/by/4.0/}{\color{linky}{CC BY 4.0}}). 
  \end{flushleft}
}

\vspace{-0.8cm}

\section{Summary}\label{summary}

\texttt{aurel} is an open-source Python package designed to
\emph{au}tomatically calculate \emph{rel}ativistic quantities. It uses
an efficient, flexible and user-friendly caching and dependency-tracking
system, ideal for managing the highly nonlinear nature of general
relativity. The package supports both symbolic and numerical
calculations. The symbolic part extends \texttt{SymPy} with additional
tensorial calculations. The numerical part computes a wide range of
tensorial quantities, such as curvature, matter kinematics and much
more, directly from any spacetime and matter data arrays using
finite-difference methods. Inputs can be either generated from
analytical expressions or imported from Numerical Relativity (NR)
simulations, with helper functions provided to read in data from
standard NR codes. Given the increasing use of NR, \texttt{aurel} offers
a timely post-processing tool to support the popularisation of this
field.

\section{Statement of need}\label{statement-of-need}

General relativity describes matter as moving according to how distances
shrink or expand; likewise, the intervals of space and time evolve
depending on the distribution of matter. Handling this dynamic ``mesh''
of distances and times requires elaborate tensor algebra that, in some
cases, can only be managed with symbolic or numerical tools. Naturally,
NR has become essential for modern astrophysics, cosmology, and
gravitational physics, most notably in the modelling of
gravitational-wave signals.

While established computational frameworks focus on solving and evolving
Einstein's field equations, with specific key diagnostics, they leave
calculations of the remaining analysis to the discretion of the
researchers. Newcomers to the field then face a substantial overhead
until they develop their own personal post-processing codes. Established
researchers also face the tedious task of handling intermediary
variables and indices when calculating new quantities. The field then
suffers from this error-prone, time-consuming process and would benefit
from an accessible, open-source, standardised framework to automate
these steps.

We therefore present \texttt{aurel}, an open-source Python package
designed to streamline relativistic calculations. It is hosted on
\href{https://github.com/robynlm/aurel}{GitHub} and is available on
\href{https://pypi.org/project/aurel/}{PyPI}. The documentation is
available through \href{https://robynlm.github.io/aurel/}{GitHub Pages}.

\section{State of the field}\label{state-of-the-field}

When looking for general relativity Python packages, there are a number
of tools that provide symbolic calculations
(\citeproc{ref-GraviPy2014}{Czaja, n.d.};
\citeproc{ref-SageManifolds2015}{Gourgoulhon, Bejger, and Mancini 2015};
\citeproc{ref-PyHole2017}{Wittig and Grover 2017};
\citeproc{ref-EinsteinPy2020}{Bapat et al. 2020};
\citeproc{ref-Pytearcat2022}{Martín and Sureda 2022};
\citeproc{ref-PyGRO2025}{Della Monica 2025};
\citeproc{ref-OGRePy2025}{Shoshany 2025};
\citeproc{ref-GREOPy2025}{Hackstein and Hackmann 2025}). Or, one may
also consider computer algebra systems
(\citeproc{ref-Maple2025}{Maplesoft 2025};
\citeproc{ref-Mathematica2025}{Wolfram Research 2025};
\citeproc{ref-xAct2025}{Martín-García et al. 2025}). However, when
non-linearities become too complex for symbolic packages, NR is used
instead.

\texttt{Einstein\ Toolkit} (\citeproc{ref-ET2012}{Löffler et al. 2012};
\citeproc{ref-ET2025}{Rizzo et al. 2025}) is a large community-driven
software whose tools enable the evolution of Einstein's field equations.
Diagnostic and further analysis calculations are typically performed on
the fly, during simulations. To study the outputs, provided by
\texttt{Carpet}, there are Python reading packages available
(\citeproc{ref-PostCactus}{Kastaun, n.d.};
\citeproc{ref-kuibit2021}{Bozzola 2021}; \citeproc{ref-scidata}{Radice,
n.d.}; \citeproc{ref-mayawaves2025}{Ferguson et al. 2025}). These extra
calculations can slow down the simulation of the spacetime evolution,
and if certain relativistic quantities are not available in
\texttt{Einstein\ Toolkit}, or in one of the post-processing packages,
then the user needs to code that up themselves.

There are a number of other well-established NR codes
(\citeproc{ref-METHOD2018}{Wright 2018};
\citeproc{ref-GRAMSES2019}{Barrera-Hinojosa and Li 2020};
\citeproc{ref-GRChombo2021}{Andrade et al. 2021};
\citeproc{ref-ExaGRyPE2024}{Zhang et al. 2025};
\citeproc{ref-MHDuet2025}{Palenzuela et al. 2025}) that also have their
own diagnostic tools. However, these are typically built-in, so going
from one code to another, to benchmark or to use their different types
of applications, requires learning the ecosystem of each.

To improve the community's versatility and limit the repeated
implementation of error-prone calculations, there is a motivation to
provide packages for computing relativistic quantities in an
NR-code-agnostic way. Especially in the post-processing sense, where all
calculations are done from a given NR spacetime and matter solution. A
couple of notable packages
(\citeproc{ref-distorted-motsfinder2018}{Pook-Kolb et al. 2019};
\citeproc{ref-BiGONLight2021}{Grasso et al. 2021}) focus on ray tracing,
or apparent-horizon finding, which are currently beyond the scope of
\texttt{aurel}. While others have more overlap
(\citeproc{ref-EBWeyl2023}{Munoz and Bruni 2023};
\citeproc{ref-EinFields2025}{Cranganore et al. 2025}) in calculating
curvature terms, they differ in scope and workflow.

Here, \texttt{aurel} innovates in its automatic design, which is easily
extendable and provides flexibility and robustness with a large and
ever-growing catalogue of relativistic quantities. A precursor to this
package was \texttt{EBWeyl} (\citeproc{ref-EBWeyl2023}{Munoz and Bruni
2023}), as it provided calculations of gravito-electromagnetic
contributions from base spacetime and matter quantities. \texttt{aurel}
now has a completely different structure (relying on the automatic
dependency resolution), provides calculations of many more terms, over
time, and has entirely new features as described in the following
section.

\section{Software Design}\label{software-design}

\texttt{aurel} provides an intuitive interface for the automatic
calculation of general relativistic quantities, either symbolically
(with \texttt{AurelCoreSymbolic}, built on \texttt{SymPy}
(\citeproc{ref-SymPy2017}{Meurer et al. 2017})) or numerically (with
\texttt{AurelCore}, which heavily utilises \texttt{numpy.einsum}
(\citeproc{ref-NumPy2020}{Harris et al. 2020}) for efficient operations
on array data structures).

Both require base quantities such as the spacetime coordinates or the
parameters of the Cartesian numerical grid, as well as the spacetime and
matter distributions (the Minkowski vacuum is otherwise assumed). These
inputs can either come from analytical expressions, with a couple of
built-in solutions available, or from output data from any NR
simulations; they just need to be passed as numpy arrays.

Specifically, for simulations run with \texttt{Carpet} in the
\texttt{Einstein\ Toolkit}, the \texttt{reading} module provides helper
functions to load and organise the 3D data. These can read the parameter
file, summarise available iterations and variables, and handle data
separated across restarts, chunks, or refinement levels for normal
\texttt{Carpet} data files or checkpoint files. To speed up repeated
data reading, \texttt{read\_data} can also split the data per iteration,
instead of per variable.

Then, once input data is provided, users can directly request a wide
range of relativistic quantities, including: spacetime; matter
(Eulerian, Lagrangian, or conserved); NR formulations; constraints;
fluid covariant kinematics; null ray expansion; 3- and 4-dimensional
curvature; gravito-electromagnetism; Weyl scalars and invariants
(including gravitational waves). To see a full list of available
quantities, see:
\href{https://robynlm.github.io/aurel/source/core.html\#descriptions-of-available-terms}{descriptions}.
Tools are also provided for spatial and spacetime covariant derivatives
and Lie derivatives. All spatial derivatives are computed by the
\texttt{FiniteDifference} class that provides 2nd, 4th, 6th and 8th
order schemes, using periodic, symmetric or one-sided boundary
conditions.

\subsection{Automatic Computational
Pathway}\label{automatic-computational-pathway}

The \texttt{aurel} automatic process composes a computational pathway at
runtime to evaluate the requested quantities. This is implemented
through a lazy-evaluation memoised property pattern, where each quantity
is defined as a method of the core class that may depend on other
quantities. This design has been chosen for its flexibility and
accessibility while remaining robust under future extensions.

Quantities are requested via a user-friendly dictionary-style access,
e.g.~\texttt{rel{[}"s\_RicciS"{]}}, which triggers the lazy memoised
check to see if this is already cached. If yes, then the result is
directly returned. If not, then the corresponding method is called,
which recursively triggers the calculation of dependencies
(e.g.~\texttt{rel{[}"s\_Ricci\_down3"{]}}). This continues until the
requested quantities can be calculated and so returned.

To avoid redundant computations, each result is cached, which builds up
a cache memory that needs to be efficiently managed. So, inspired by
Python's garbage collection, \texttt{aurel} uses an intelligent eviction
policy that tracks memory footprint, evaluation counts, and last-access
times. When the configurable thresholds are exceeded, the older and
heavier cached quantities are removed, while safeguarding protected base
quantities. Throughout this process, \texttt{aurel} keeps the user
informed on progress by providing verbose updates on the computation and
caching workflow.

\subsection{Time dependence}\label{time-dependence}

All calculations within the \texttt{AurelCore} class are evaluated at a
single fixed time, corresponding to one slice in time, so that
individual time steps can be treated independently. For multiple time
steps, an \texttt{AurelCore} object needs to be created and the
requested quantities collected for each.

To streamline this process, \texttt{aurel} provides the
\texttt{over\_time} function to do exactly this, and also compute
summary statistics over the grid domain (e.g., max/min) at each time
step. By design, it is easily extensible, so \texttt{over\_time} also
accepts custom functions of new relativistic quantities and summary
statistics. This makes \texttt{aurel} versatile, supporting an infinite
number of ways to view a problem and develop diagnostic tools.

\section{Research Impact Statement}\label{research-impact-statement}

\texttt{aurel} is a specialist tool for general relativity researchers
and streamlines numerical relativists' post-processing workflow. Through
conference interactions and collaborations involving the authors, this
package has gradually been disseminated to individual researchers who
appreciate the effortless integration, satisfying dependency resolution
and substantial reduction to post-processing overhead. Indeed, in
ongoing studies involving NR simulations of primordial black hole
formation, \texttt{aurel} has increased capacity and redirected
repetitive and error-prone development efforts towards exploring a
broader range of simulated scenarios. Additionally, for master students,
the straightforward and transparent design has provided an easy gateway
for them to analyse NR simulations and so quickly get results within the
duration of their projects. Going forward, awareness of this code will
build upon publication, reaching a wider audience and supporting the
popularisation of NR.

\section{AI usage disclosure}\label{ai-usage-disclosure}

GitHub Copilot Claude Sonnet 4 was used for the development and
documentation of this package. Autocompletion suggestions were accepted
via the VSCode Copilot plugin, and upon the developer's request, edits
and code snippets were generated via the large language model's user
interface. The most significant AI contributions came in drafting the
docstrings and scaffolding the test suite, both of which are essential
for the accessibility and robustness of this package. Each and every
suggestion or contribution was meticulously reviewed and adjusted before
being included by the authors, who made all core design decisions and
innovated the original structural concept. Finally, this paper was
prepared without the use of generative language models, solely with
grammar checkers.

\section{Acknowledgements}\label{acknowledgements}

We thank Nat Kemp for being one of the first testers of \texttt{aurel}.
We thank Ian Hawke for support and suggestions.

RM and WR are supported by an STFC grant ST/X001040/1. CB is supported
by STFC grants ST/X001040/1 and ST/X000796/1.

\phantomsection\label{refs}

\end{document}